\documentclass{Interspeech2024}
\usepackage{multirow, bbding, makecell} 
\usepackage{setspace}
\usepackage{xurl}
\usepackage{hyperref}

\interspeechcameraready

\title{BS-PLCNet 2: Two-stage Band-split Packet Loss Concealment Network\\ with Intra-model Knowledge Distillation}

\name[affiliation={1,2}]{Zihan}{Zhang}
\name[affiliation={2}]{Xianjun}{Xia}
\name[affiliation={2}]{Chuanzeng}{Huang}
\name[affiliation={2}]{Yijian}{Xiao}
\name[affiliation={1*}]{Lei}{Xie}

\address{
  $^1$Audio, Speech and Language Processing Group (ASLP@NPU), School of Computer Science, \\ Northwestern Polytechnical University, Xi'an, China\\
  $^2$ByteDance, China }
\email{zhzhang@mail.nwpu.edu.cn, lxie@nwpu.edu.cn\thanks{$^*$: Corresponding author.}}


\begin{document}
\maketitle
\begin{abstract}
Audio packet loss is an inevitable problem in real-time speech communication. 
A band-split packet loss concealment network (BS-PLCNet) targeting full-band signals was recently proposed. Although it performs superiorly in the ICASSP 2024 PLC Challenge, BS-PLCNet is a large model with high computational complexity of 8.95G FLOPS. This paper presents its updated version, BS-PLCNet 2, to reduce computational complexity and improve performance further. Specifically, to compensate for the missing future information, in the wide-band module, we design a dual-path encoder structure (with non-causal and causal path) and leverage an intra-model knowledge distillation strategy to distill the future information from the non-causal teacher to the casual student. Moreover, we introduce a lightweight post-processing module after packet loss restoration to recover speech distortions and remove residual noise in the audio signal. With only 40\% of original parameters in BS-PLCNet, BS-PLCNet 2  brings 0.18 PLCMOS improvement on the ICASSP 2024 PLC challenge blind set, achieving state-of-the-art performance on this dataset.

\end{abstract}

\section{Introduction}
In the past few years, the advancement of the internet and technologies coupled with the pandemic have gradually made voice over internet phone (VoIP) technology an essential part of our daily life\cite{nguyen2023improving}.
However, due to various factors, such as network congestion, delay and jitter, hardware failures, audio data packets may get lost during transmission, which degrades the speech quality~\cite{lin2021time}.
Packet loss concealment (PLC) technology is applied to restore the missing content, thus improving the speech quality and intelligibility.

In contemporary speech codecs, traditional interpolation~\cite{perkins1998survey} or Hidden Markov Model (HMM) based approaches~\cite{borgstrom2010efficient} are typically employed for PLC. 
While these approaches only work well in scenarios with short-time lost, and can not handle conditions of high packet loss rate and long-gap lost, leading to a significant degrade in speech quality. In recent years, with the development of deep learning, methods based on deep neural networks (DNN) have shown great potential in PLC. The early deep PLC approach~\cite{lee2015packet} used log-spectra to train a Time-Frequency (T-F) domain DNN model as a nonlinear regression function for PLC. As T-F domain approaches need to recover the phase information to reconstruct the waveform, the performance may be limited~\cite{li2022end}. More recent approaches show a preference for using time-domain models~\cite{lin2021time,li2022end, westhausen2022tplcnet,valin2022real}. Lin et al.~\cite{lin2021time} propose a time-domain convolutional recurrent network (CRN) for online packet loss concealment. Westhausen et al.~\cite{westhausen2022tplcnet} predict lost frames from a short context buffer in a sequence-to-one (seq2one) fashion by only performing the inference. Valin et al.~\cite{valin2022real} implement PLC in the feature domain. They use a predictive conditioning model to predict the acoustic features and a generative vocoder to synthesis speech. Li et al.~\cite{li2022end} propose a time domain generative adversarial network (GAN) structure, using discriminator and multi-loss to ensure both speech quality and automatic speech recognition (ASR) compatibility. 

Recently, audio transmission has transitioned to full-band signals (48KHz). However, due to the latency constraint and high computational cost, the aforementioned approaches work only at wide-band signals (16kHz). 
The latest ICASSP 2024 PLC Challenge promotes full-band real-time PLC and focuses on more challenging long-gap packet lost. Among all the submissions, BS-PLCNet~\cite{zhang2024bs} with a band-split gated convolutional recurrent neural network leads to superior performance on the blind test set, showing the effectiveness of T-F domain models. However, it has a high computational complexity of 8.95G FLOPS. Additionally, the common encoder-decoder structure in BS-PLCNet fails to learn a robust feature structure as future information is missing.

In this paper, we propose BS-PLCNet 2 to further improve the PLC performance and reduce the computational complexity. There are three contributions in this paper: 
\vspace{-1pt}
\begin{itemize}
    \item We optimize the network structure of BS-PLCNet for superior feature extraction and replace the original convolution operation with depth-wise separable convolution~\cite{luo2019conv}.
    \item Knowledge distillation strategy is adapted. We upgrade the model with intra-model knowledge distillation~\cite{ning2023dualvc}, which allows us to obtain the final model without training the teacher and student model separately.    
    \item We incorporate a post-processing module and utilize a two-stage training strategy to recover speech distortions and remove residual noise in the audio signal.
\end{itemize}
\vspace{-1pt}
Experimental results show the proposed BS-PLCNet 2 outperforms BS-PLCNet with a 0.18 OVRL PLCMOS~\cite{diener2023plcmos} improvement on the ICASSP 2024 PLC Challenge blind set and achieves state-of-the-art (SOTA) results. Remarkably, the non-causal model only has 38.1\% and 40\% of original computational cost and number of parameters, respectively.

\section{Proposed Method}
\begin{figure*}[t]
    \centering
    \includegraphics[scale=0.60]{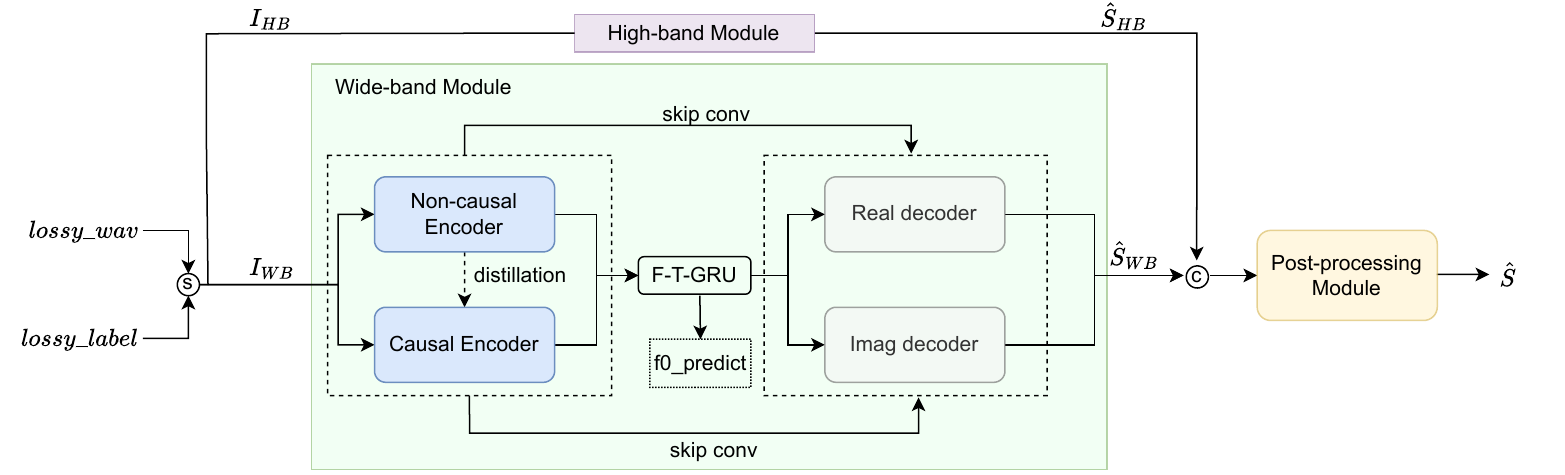}
    \caption{The proposed BS-PLCNet 2.}
    \label{fig:fig1}
\end{figure*}
\begin{figure*}[t]
    \centering
    \includegraphics[scale=0.6]{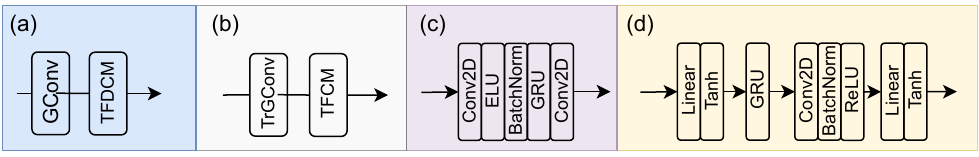}
    \caption{Details of the encoder layer (a), the decoder layer (b), the high-band moudle (c) and the post-processing module (d).}
    \label{fig:fig_details}
\vspace{-6pt}
\end{figure*}
\subsection{Problem formulation}
Let $s \in \mathbb{R}^{ L} $ be an original speech, where $L$ is the length of the signal. Considering the potential additive noise $v \in \mathbb{R}^{ L} $, the transmission signal $x$ can be represented as $x=s + v$.  For audio transmission, the continuous audio $x$ is often split into short-time frames with the frame length $N$. The $k$th speech frame can be defined as:
\begin{equation}
\label{eq1} x_{k}(n)=x(N\cdot(k-1)+n ), n=0,1,...,N-1 ,
\end{equation}
where $n$ denotes the time sample index. The packet loss flag is defined as $F$. If packet loss flag is 1, the frame is lost, and all samples in this frame are set to zero. The received signal can be defined as:
\begin{equation}
\label{eq2} 
\begin{cases}
x_k = s_k + v_k, & F_k=0
\\
x_k = [0,0,...,0], & F_k=1
\end{cases}
\end{equation}
In the frequency domain, Eq.~\ref{eq2} can be formulated as:
\begin{equation}
\label{eq3} 
\begin{cases}
X(t,f) = S(t,f) + V(t,f), & F_t=0
\\
X(t,f) = 0, & F_t=1
\end{cases}
\end{equation}
where $t$ and $f$ are the frame index and the frequency index, respectively. Given the 50\% overlap used during short-time Fourier transform (STFT), the correspondence between the transmission frame index $k$ and the STFT frame $t$ is: 
$F_k=1, F_{t=2k}=F_{t=2k+1}=1$.

\subsection{Intra-model knowledge distillation}
Given that the causal PLC model only utilizes historical information and cannot look ahead to capture future information, which has shown to be effective in~\cite{diener2022interspeech}. We propose to use dual-path convolution in couples with intra-model knowledge distillation to address this issue. 
Similar structures have yielded notable results in auto speech recognition (ASR)~\cite{liang2023fast} and voice conversion~\cite{ning2023dualvc}, which also have a significant perception of future information. 

As shown in Fig.~\ref{fig:fig1}, the dual-path convolution is based on depth-wise separable convolution~\cite{luo2019conv} and contains both causal and non-causal paths. 
In parallel depth convolution layers, the padding mode of the depth convolution is modified to implement causal and non-causal paths. Meanwhile, the parameters of the parallel depth convolution layers are not shared. Specifically, in gated convolutional module~\cite{zhang2022multi} and the time-frequency dilated convolutional module (TFDCM)~\cite{sun2023multi}, all the convolutions in time dimension have been replaced with dual-path convolutions. 

Unlike the typical knowledge distillation in teacher-student networks~\cite{li2014learning,hinton2015distilling, chebotar2016distilling}, We only design dual paths in the convolution layer of the encoder, while other layers share parameters.
Consequently, both causal and non-causal inputs can be simultaneously received by the decoder and the bottleneck layer during training, which reduces the difficulty of knowledge transfer. This approach also allows us to obtain the final model in one single step, avoiding the need first to train the teacher model and then distill the student model. By using the intra-model knowledge distillation strategy, the training cost is reduced significantly.

During the training process, the input data stream is directed through two distinct paths. To maximize the similarity between the causal encoder output and the non-causal encoder output, we calculate the distillation loss between the causal encoder's output and the non-causal encoder's output. Dual-path convolution not only allows for the concurrent training of both causal and non-causal models in one step, but it also employs the non-causal model to guide the causal model during training, thus improving the performance of the causal model by using the future information.

\subsection{Network optimization and two-stage post-processing}
\label{sec2.3}
Similar to BS-PLCNet~\cite{zhang2024bs}, we retain the Generative Adversarial Network (GAN) architecture. For the generator, we have the following improvements. First, we replace all convolution layers with depth-wise separable convolution layers to reduce the computational complexity of BS-PLCNet. Secondly, the layer number within the TFDCM module is modified. The proposed layer number is dynamically adaptive in accordance with the network depth rather than using fixed 4 layers in~\cite{zhang2024bs}. The dynamic way to set the layer number reduces the receptive field and allows a stronger focus on local information, which is achieved as neural network layers deepen and the feature dimensions increase. Finally, we replace the bottleneck layer from three F-T-LSTM~\cite{zhang21ia_interspeech} layers to a single F-T-GRU layer~\cite{chen2023progressive}. Stacking multiple dual-path layers for the same sequence can lead to limited performance~\cite{tong2024scnet}.

Motivated by the multi-stage network for speech enhancement (SE), we also introduce a post-processing module to recover the potential distortions. Potential noise is also addressed at this stage. Considering the overall computational complexity increase, the post-processing module only processes the 0-8kHz frequency bands with a simple GRU-based structure, which consists of a Linear layer activated by Tanh, a GRU layer, a 2D convolution (Conv2d) layer, and a batchnorm layer followed by another Linear layer activated by Tanh. 

During training, we applied a two-stage training strategy. In the first stage, we utilize multi-frequency discriminators (MFD)~\cite{tian2020tfgan} and multi-period discriminators (MPD)~\cite{kong2020hifi} to train the first stage of the BS-PLCNet 2. During the first training stage, the model focuses on recovering lost audio samples by leveraging the received audio signal and packet loss information. In the second stage, we use MetricGAN~\cite{fu2021metricgan+} to train the post-processing module. The parameters of the first stage module are frozen during the second stage of training. The second stage input is a mixture of the first stage output and additive noise.
\subsection{Loss function}
For the first training stage, we utilize speech loss $\mathcal{L}_\text{speech}$ and GAN loss. The speech loss is composed of a power-law compressed phase aware (PLCPA) loss $\mathcal{L}_\text{PLCPA}$~\cite{zhang2022multi} and time-domain MAE loss $\mathcal{L}_\text{MAE}$.
{
\begin{equation}
\mathcal{L}_\text{PLCPA}=\mathcal{L}_\text{mag} + \mathcal{L}_\text{pha},
\vspace{-4pt}
\end{equation}
}
{
\begin{equation}
\mathcal{L}_\text{MAE}=\sum_{i=0}^{N} |s_{i} - \hat{s}_{i}|, 
\end{equation}
}
{
\begin{equation}
\mathcal{L}_\text{speech}=\mathcal{L}_\text{PLCPA} + \mathcal{L}_\text{MAE}.
\end{equation}
}
For the GAN loss, we use the least-square GAN (LSGAN)~\cite{mao2017least} style training objectives, which can be defined as:
{
\begin{equation}
\mathcal{L}_\text{D}=(1-(D(s)))^{2}+D(G(x))^{2},
\end{equation}
}
{
\begin{equation}
\mathcal{L}_\text{G}=(1-D(G(x)))^{2}.
\end{equation}
}
The MAE based f0 prediction loss $\mathcal{L}_\text{f0}$ is still retained as:
{
\begin{equation}
\mathcal{L}_\text{f0}=\sum_{i=0}^{T} |f_{i} - \hat{f}_{i}|
\end{equation}
}
where $T$ represents the frame index.
The overall multi-loss for the first training stage is:
{
\begin{equation}
\mathcal{L}_\text{first}=\mathcal{L}_\text{PLCPA} + \mathcal{L}_\text{MAE} + \alpha \cdot \mathcal{L}_\text{f0} + \mathcal{L}_\text{G},
\end{equation}
}
where $\alpha$ equals $1e^{-1}$, and for the second training stage, we only use speech loss to train the post-processing module.

\section{Experiments}

\subsection{Datasets}
For the training data, we use train-clean-100 and train-clean-360 from Librispeech~\cite{panayotov2015librispeech} and the 5th DNS Challenge dataset\footnote{\url{https://github.com/microsoft/DNS-Challenge}} as the original clean speech, from which the training lossy signals are simulated. The training set contains nearly 800 hours data in total.

\subsection{Pack loss rate configuration}
\begin{figure}[t]

\begin{minipage}{0.48\linewidth}
    \includegraphics[scale=0.6]{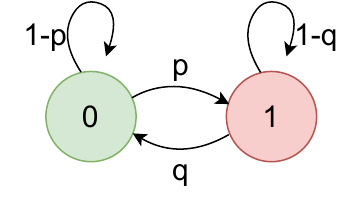}

\end{minipage}
\hfill
\begin{minipage}{0.48\linewidth}
    \begin{tabular}{@{}ccc@{}}
    \toprule
    p  & q & Expected loss(\%) \\ \midrule
    0.1 &  0.9       & 10    \\
    0.1 &   0.5     & 16.7    \\
    0.5 &  0.9     & 35.7    \\
    0.9    &  0.1     & 50     \\ 
     \bottomrule
    \end{tabular}
\end{minipage}
\caption{The Gilbert-Elliott Model.}
\label{fig:fig2}
\vspace{-8pt}
\end{figure}

We use the Gilbert-Elliott model~\cite{mushkin1989capacity} to model the conditional consecutive packet loss. As shown in Fig.~\ref{fig:fig2}, the Gilbert-Elliott model uses a two-state first-order Markov chain to describe the packet loss status of the current frame and the next frame, where $0$ indicates no packet loss, and $1$ denotes the existence of packet loss. If the current frame transmits normally, The next frame has a $p$ probability of being lost. If the current frame is lost, the probability of the next frame not being lost is $q$. By setting $p$ and $q$, the expected packet loss rate $r$ can be controlled as:
{
\begin{equation}
r=p/(p+q).
\label{eql}
\end{equation}
}

The packet loss rate is set to not exceed 50\%. During training, we first randomly select two transition probabilities p and q, ranging from 10\% to 90\%. Then we check whether the expected packet loss rate is below 50\% according to Eq.~\ref{eql}. If the packet loss rate condition is not satisfied, then $p$ and $q$ are reset. Note that $r$ is the expected packet loss rate, actual packet loss rate for individual sentence may vary.

For the first training stage, we only simulate packet loss. Noise is introduced in the second training stage. We use the 5th DNS Challenge noise dataset to simulate the noisy mixtures. The training data are generated and segmented into 8s chunks in each batch with signal-to-noise ratio (SNR) ranging from 10 to 20dB.

\begin{table*}[htb!]
\centering
\caption{The objective scores on the 2024 blind test set. Here, the ``Structure optimization", the ``Dual conv" and the ``Two-stage" strategies are for optimizing the complexity, utilization of future information and post-processing, respectively.}

\label{tab:exp1}
\begin{tabular}{@{}clccccccccccc@{}}
\toprule 
\makecell[c]{System} &Model   & Para.(M) & FLOPs(G) & \makecell[c]{Structure \\ optimization} & \makecell[c]{Dual\\conv}  & \makecell[c]{Two\\stage}& Causal & PLCMOS & SIG & BAK & OVRL \\ \midrule
S0 &lossy  & -  & -& & & & & 2.69  & 3.17   & 3.92   & 2.89  \\
S1& Team 1024K~\cite{li2024multi}    & -         & -  & & & & \checkmark      & 4.16       & -      & -  &  -   \\ \midrule
S2& BS-PLCNet~\cite{zhang2024bs} &  6.50         & 8.95  & \texttimes &  \texttimes & \texttimes & \checkmark     & 4.02       &  3.62    &   4.09  & 3.35  \\
S3&BS-PLCNet 2    & 2.01         & 3.30  & \checkmark & \texttimes & \texttimes  & \checkmark      & 4.05       &  3.63     & 4.09   &   3.35     \\ 
S4& BS-PLCNet 2   & 2.01         & 3.30   & \checkmark & \checkmark & \texttimes & \checkmark     & 4.14      & \textbf{3.64}     & 4.10 & \textbf{3.37}       \\ 
S5& BS-PLCNet 2   & 2.62         & 3.41   & \checkmark & \checkmark & \checkmark &  \checkmark     & \textbf{4.20}  & 3.61   & \textbf{4.15}  & \textbf{3.37}       \\ \midrule
S6&BS-PLCNet 2-NC   & 2.62         & 3.41   & \checkmark & \checkmark & \checkmark  & \texttimes     & 4.30    &   3.66   & 4.14    &   3.40     \\ \bottomrule
\end{tabular}
\end{table*}

\begin{table}[htb!]
\centering
\caption{The objective scores on the 2022 blind test set.}
\label{tab:exp2}
\begin{tabular}{@{}lccccc@{}}
\toprule
Model  & PLCMOS & OVRL & PESQ & STOI \\ \midrule
lossy & 2.90  &  2.56  & 2.37  & 0.84  \\
Team Kuaishou~\cite{li2022end}           & 4.27       & -      & 3.01  &  -   \\
LPCNet~\cite{valin2022real}    & 3.74      &  3.09    &   2.93  & 0.90  \\
PLCNet~\cite{liu2022plcnet}      & 3.83 & -   & 2.37  & 0.87       \\
BS-PLCNet~\cite{zhang2024bs}         &  4.29      &  3.20     &  3.04  & 0.90       \\ 
BS-PLCNet 2     & \textbf{4.36}      & \textbf{3.25}     & \textbf{3.22} & \textbf{0.92}   \\ 
 \bottomrule
\end{tabular}
\vspace{-4pt}
\end{table}
\subsection{Experimental setup}
The proposed BS-PLCNet 2 works in T-F domain, we apply the Hanning window for each frame, the frame shift and FFT points are set to 20ms, 10ms and 960, respectively. We train the model with the Adam optimizer~\cite{kingma2014adam}. The initial learning rate is set to $1e^{-3}$ and $1e^{-4}$ In the first and second training stages, respectively. The learning rate is halved as the validation loss does not decrease in two epochs. We apply a maximum L2 norm of 5 for gradient clipping. Each stage is trained for 30 epochs.
For the wide-band module, both the encoder and decoder have 4 layers, each with 80 channels, except for the first layer of the encoder and the last layer of the decoder. The kernel size and the strides are set to (2, 3) and (1, 2) in time and frequency axes, respectively.

In the TFDCM module, the stacked $n$ convolutional layers are adopted with a dilation rate of [$2^0$, $2^1$,..., $2^{n-1}$] in both the time and frequency axis. From the first layer to the fourth layer of the encoder, $n$ is respectively set as [4,4,3,2]. The stride and kernel size of DConv in TFDCM are set to (1, 1) and (3, 3). The TFCM module is the same as that in~\cite{ju2023tea}. The F-T-GRU layer is a single GRU layer with 128 hidden states. 
For the high-band module, the first Conv2D has an output channel of 128, and the GRU layer has a hidden state of 128. 

\subsection{Metrics}
This paper evaluates different systems on the blind test sets of ICASSP 2024 and Interspeech 2022 PLC Challenges. Note that labels for the former are unavailable, while labels for the latter are publicly available. On the ICASSP 2024 PLC Challenge blind test, PLCMOS~\cite{diener2023plcmos} and DNSMOS~\cite{reddy2022dnsmos} are used, which are both neural networks trained to estimate the ratings human raters would assign to an audio file. Specifically, DNSMOS consists of speech quality (SIG), background noise quality (BAK), and the overall quality (OVRL) of the audio. 
On Interspeech 2022  PLC Challenge blind test, PLCMOS, DNNMOS, PESQ~\cite{rix2001perceptual} and STOI~\cite{taal2011algorithm} are used to evaluate different systems. It is worth noting that the released PLCMOS version from Microsoft is different for these two challenges.

\subsection{Results and analysis}
In Table~\ref{tab:exp1}, we perform ablation experiments for the proposed method on the blind test set of ICASSP 2024 PLC Challenge and compare it with two other systems, Team 1024K~\cite{li2024multi} and BS-PLCNet~\cite{zhang2024bs}, which tied for the 1st place in the challenge.

To verify the effectiveness of optimizing the network structure in Section~\ref{sec2.3}, we perform the system comparison between BS-PLCNet (S2) and BS-PLCNet 2 (S3) as shown in Table 1. The new network structure achieves comparable PLCMOS and DNSMOS with a substantial reduction in both computational complexity and the number of parameters. 

To show that utilization of future information through knowledge distillation in a casual way is beneficial to the PLC system, BS-PLCNet 2 with dual-conv (S4) is compared with S3. As can be seen from Table~\ref{tab:exp1}, the use of intra-model knowledge distillation significantly improves PLCMOS from 4.05 to 4.14. This observation demonstrates that under the guidance of the non-causal model, the causal model achieves superior performance without using the feature information. It is worth noting that since the dual-path convolution only executes one branch during inference, the computational complexity and parameters remain the same. 

To investigate the importance of our proposed post-processing module and two-stage training strategy,  BS-PLCNet 2 combining dual-conv and two-stage post-processing strategy (S5) is compared with system S4. The post-processing module we added has improved both PLCMOS and DNSMOS, especially the BAK scores, while only leading to an increase of 0.11 GFLOPs in computational complexity. The use of the post-processing module and two-stage training strategy enables the model to handle potential distortions and noise, thus improving the quality of the generated speech.

To show the system upper bound by using the future information in a non-casual way, BS-PLCNet 2-NC (S6) shows the result of the non-causal branch in the dual-path convolution. Due to its ability to access future information, the non-causal model still has an advantage in PLC. 

\urldef{\myurl}\url{https://Interspeech-BSPLCNet2.github.io/BS-PLCNet2}
In Table~\ref{tab:exp2}, we also use the Interspeech 2022 blind test set to compare our system with the top three systems from the challenge, which are Team Kuaishou~\cite{li2022end}, LPCNet~\cite{valin2022real}, and PLCNet~\cite{liu2022plcnet}. The results show that the proposed BS-PLCNet 2 continues to retain its superiority when compared with more systems. The two-stage band-split PLCNet with intra-model knowledge distillation outperforms other methods in all metrics. Especially in PESQ, compared to the top 1 team Kuaishou, the proposed method brings 0.21 PESQ performance improvement. The results on both the ICASSP 2024 blind test set and the Interspeech 2022 blind test set show that the proposed BS-PLCNet 2 achieves SOTA. Our samples are available \footnote{\myurl}.

\section{Conclusion}
In this paper, we propose a novel two-stage band-split packet loss concealment network – BS-PLCNet 2, which is an upgrade of BS-PLCNet. Benefiting from the intra-model knowledge distillation, BS-PLCNet 2 achieves the state-of-the-art PLCMOS performance in the ICASSP 2024 PLC Challenge blind test set with only 38.1\% FLOPs of BS-PLCNet in computation complexity. Additionally, both causal and non-causal models can be obtained simultaneously within a single training process. In the future, we will particularly consider the efficient use of temporal information, also a combination in the time-frequency domain to further improve speech quality.
\bibliographystyle{IEEEtran}
\bibliography{mybib}

\end{document}